\documentclass[11pt]{article}
\usepackage{jheppub}
\usepackage{braket} 

\usepackage{tikz}
\usetikzlibrary{intersections, calc,positioning,decorations.pathreplacing,decorations.pathmorphing,arrows,arrows.meta,patterns}

\usepackage{amsthm}

\def\tr{{\rm tr}}
\def\Tr{{\rm Tr}}

\title{$3d$ One-form Mixed Anomaly and Entanglement Entropy}

\author{Yang Zhou}

\affiliation{Department of Physics and Center for Field Theory and Particle Physics,\\
Fudan University, Shanghai 200433, China\\}

\abstract{
We study mixed anomaly between $G_1$ and $G_2$ of one-form finite symmetry $G_1\times G_2$ in $3d$ Chern-Simons theories. We assign a quantum entanglement structure to two linked $G$-symmetry lines (Wilson loops) and compute the entanglement entropy $S[G]$. We find a measure of the mixed anomaly by computing $S[G_1\times G_2]-S[G_1]-S[G_2]$.
}

\preprint{}

\begin{document}
\maketitle

\section{Introduction}
The goal of this paper is to propose a measure of the mixed anomaly between $G_1$ and $G_2$ where both of them are $3d$ $1$-form global symmetries. 't Hooft anomaly for a global symmetry $G$ is an obstruction for gauging $G$~\cite{tHooft:1979rat}. If $G$ is a connected Lie group, its 't Hooft anomaly is tightly constrained by Wess-Zumino consistency conditions~\cite{Wess:1971yu}. This implies the anomaly inflow mechanism. Namely possible anomalies for $G$ in $d$ spacetime dimensions are classified by Chern-Simons actions in dimension $d+1$. When $G$ is a finite group, Wess-Zumino constraints do not apply but the inflow mechanism still works. Recently there has been a lot of interest to study quantum field theories with global discrete symmetries~\cite{Sato:2006de,Ryu:2012he,Hung:2018rhg,Han:2017hdv,Chen:2011pg,Gaiotto:2014kfa,Gaiotto:2017yup,Gaiotto:2017tne,Seiberg:2016gmd,Cordova:2017kue,Cordova:2018cvg,Hsin:2018vcg,Tachikawa:2016nmo,Tanizaki:2017bam,Yamaguchi:2018xse,Armoni:2018bga,Kitano:2017jng,Wang:2018edf,Guo:2018vij,Yao:2018kel,Tanizaki:2018xto,Numasawa:2017crf}. When $G$ is a product form $G=G_1\times G_2$, it is possible that $G$ has a nontrivial 't Hooft anomaly while both $G_1$ and $G_2$ have trivial 't Hooft anomaly, which means that $G_1$ or $G_2$ can be gauged, but not the whole $G$~\cite{Kapustin:2014lwa,Vishwanath:2012tq}. In this case there is a mixed anomaly between $G_1$ and $G_2$. Recently the ordinary symmetries ($0$-form) have been generalized to higher form. Under an ordinary internal global symmetry the charged operators are point-like. A $p$-form symmetry is a global symmetry under which the charged operators are of spacetime dimension $p$ and the topological symmetry defects have co-dimension $p+1$ in spacetime~\cite{Gaiotto:2014kfa}.

In $3d$ Chern-Simons theory, $1$-form symmetry lines and the charged operators have the same spacetime dimension one. They are realized as Wilson lines. As examples, $U(1)$ Chern-Simons theory at level $k$ and $SU(N)$ Chern-Simons theory at level $k$ both have 1-form symmetries. The former has 1-form symmetry $Z_k$ and the latter has 1-form symmetry $Z_N$ which coincides with the center of the gauge group. The 't Hooft anomaly of $3d$ $1$-form symmetry $G$ can be detected by examining whether the symmetry lines of $G$ are charged under themselves. To illustrate the associated anomaly, one can label the symmetry lines by $U_g$ where $g$ is a group element $g\in G$. The group multiplication corresponds to the fusion of the symmetry lines:
\begin{equation}
U_{gg'} = U_g\times U_{g'}\ .
\end{equation}
In the case of $Z_N$ 1-form symmetry, the generator can be chosen as $U$ such that
$U^N=1$. Given any line operator $V$, the $Z_N$ charge $q(V)$ under $U$ can be measured by
\begin{equation}\label{symact}
U^s(a) V(b) = e^{2\pi i q(V)s\over N} V(b)\ ,
\end{equation} where $a$ is a circle around $b$. This gives the charge of $U$ under itself,
\begin{equation}\label{symano}
U(a) U(b) = e^{2\pi i q(U)\over N} U(b)\ .
\end{equation}
A nontrivial phase in (\ref{symano}) shows a 't Hooft anomaly of $Z_N$, therefore $q(U)=0,1,\cdots,N-1$ may be used to classify different anomalies.~\footnote{Through this paper we do not consider fermionic lines appearing in spin TQFT. We refer to~\cite{Hsin:2018vcg} for related discussion.} In particular, only when $\gcd(q(U),N)=1$, the $Z_N$ symmetry is fully anomalous, otherwise the fully anomalous symmetry among the symmetry lines becomes $Z_{N/\gcd(q(U),N)}$. Here we define ``fully anomalous $G$'' by requiring that there is no subgroup of $G$, which is anomaly free and also does not have mixed anomaly with other symmetries. In~\cite{Hsin:2018vcg} Hsin, Lam and Seiberg use the spin of the generating line $h[U]$ to classify $Z_N$ 1-form symmetries.~\footnote{I also learned this from Zohar Komargodski's lectures.}
Since $q(U)$ is determined by $h[U]$, it can be used to classify different $Z_N$ $1$-form symmetries as well.

If the 1-form symmetry is a product form $G=G_1\times G_2$, a mixed anomaly between $G_1$ and $G_2$ exists when the symmetry lines of one group are charged under the other. For simplicity let us consider $G_1=Z_M$ and $G_2=Z_N$. Write equation (\ref{symano}) for $G_1$ and $G_2$,
\begin{equation}\label{anomalyeqG1G2}
U_1(a) U_1(b) = e^{2\pi i q_1\over M} U_1(b)\ ,\quad U_2(a) U_2(b) = e^{2\pi i q_2\over N} U_2(b)\ .
\end{equation} There are also crossing equations
\begin{equation}\label{anomalyeqG12}
U_1(a) U_2(b) = e^{2\pi i p_2\over M} U_2(b)\ ,\quad U_2(a) U_1(b) = e^{2\pi i p_1\over N} U_1(b)\ .
\end{equation}
One can measure the expectation values of both sides in (\ref{anomalyeqG1G2}) and (\ref{anomalyeqG12}) on a 3-manifold such as $S^3$. The VEV of linked Wilson loops in $S^3$ is the modular S matrix, which is symmetric. This implies~\footnote{The vacuum expectation values (VEV) of a single symmetry line (along a circle) are normalized to be the same 
\begin{equation}
\langle U_1\rangle = \langle U_2\rangle\ .
\end{equation}}
\begin{equation}\label{mixc}
{p_2\over M} = {p_1\over N}~\text{mod}~1.
\end{equation}
Equation (\ref{mixc}) constrains the possible mixed anomalies between $Z_M$ and $Z_N$. Moreover, the symmetry property of modular S matrix implies ($\lambda=e^{2\pi i p_2\over M}=e^{2\pi i p_1\over N}$)
\begin{equation}\label{lambda}
U^s_1(a) U^{s'}_2(b) = \lambda^{ss'} U_2(b)\ ,\quad U^{s'}_2(a) U^s_1(b) = \lambda^{ss'} U_1(b)\ ,
\end{equation} where $s=0,1,\cdots, M-1$ and $s'=0,1,\cdots, N-1$. Therefore two integers $q_1, q_2$ and a phase $\lambda$ together specify all the braiding phases between the symmetry lines of $Z_M\times Z_N$.

For a given anyon system described by some topological field theory, how to detect the 't Hooft anomaly for a 1-form symmetry? One can in principle check the explicit braiding phases, but this can be tedious once the symmetry becomes large. To obtain a measure of the anomalous symmetries, the authors in \cite{Hung:2018rhg} turn to consider the entanglement entropy coming from the braiding between Wilson loops (anyons). This happens as follows. The modular S matrix can be identified as the wave function of a state on two linked tori~\cite{Salton:2016qpp,Balasubramanian:2016sro,Dwivedi:2017rnj}. The Hilbert space is a tensor product of two torus Hilbert spaces. One can also view the modular S matrix as the wave function of a pair of entangled bits, each staying in a torus Hilbert space. Restricted to the symmetry lines, the wave function is a truncated modular S matrix. It has been proposed~\cite{Hung:2018rhg} that the entanglement entropy for the truncated modular S matrix can be used as a measure of the anomalous symmetries~\footnote{See (\ref{entropyanomaly}) for details.}
\begin{equation}\label{anomalymeasure}
S=\log D\ ,
\end{equation} where $S$ is the entanglement entropy and $D$ is the order of the anomalous group. This has passed the tests of all known examples. By this mean, zero entropy of a symmetry $G$ will reflect that $G$ does not have 't Hooft anomaly. While a finite entropy $\log D$ reflects that there is an anomalous symmetry with order $D$. However this is not enough to show the origin and the detail structure of the finite 't Hooft anomaly. In particular we do not know whether this anomaly comes from a mixed anomaly or not.

In this paper, we support the idea of measuring the anomaly using entropy by moving a step forward. We show that an entropy measure
\begin{equation}\label{mixmeasure}
\Delta S := S[G_1\times G_2]-S[G_1]-S[G_2]
\end{equation} is a good measure of the mixed anomaly between $G_1$ and $G_2$. This formula (\ref{mixmeasure}) implies that the mixed anomaly corresponds to the {\it residual entropy.} We check our proposal for known examples.

$1$-form symmetry in $3d$ bulk topological field theory becomes $0$-form symmetry in $2d$ boundary theory, because $1$-dimensional symmetry lines become the topological defect lines of $0$-form symmetry in $2d$~\cite{Chang:2018iay,Bhardwaj:2017xup}. This can be seen from the $2d$ counterpart of $3d$ Chern-Simons theory,  the Wess-Zumino-Witten models. The $1$-form anomaly of Chern-Simons theory therefore becomes $0$-form anomaly in Wess-Zumino-Witten models. The 't Hooft anomaly of a $0$-form $G$-symmetry in bosonic $2d$ theory is classified by the cohomology group $H^3(G,U(1))$~\cite{Dijkgraaf:1989pz}. Our truncated modular S matrix approach (\ref{anomalymeasure})(\ref{mixmeasure}) to detect the 't Hooft anomaly and mixed anomaly can also be used in diagonal rational conformal field theory (RCFT) in a straight manner.~\footnote{To our best knowledge our criterions give correct anomaly results for all the known $2d$ RCFT examples with discrete internal global symmetry.} This is because there are chiral vertex primaries which correspond to topological defect lines associated to a internal global symmetry~\cite{Moore:1988qv,Moore:1989yh,Elitzur:1989nr}.

In~\cite{Han:2017hdv}, it has been proposed that one can use $G$-invariant boundary state condition in CFT$_2$ to justify whether there is a 't Hooft anomaly or not, namely the existence of a $G$-invariant boundary state will imply that $G$ is 't Hooft anomaly free. By this mean,~\cite{Numasawa:2017crf} computed the anomaly free condition (conditions on the level $k$) for the center symmetry of Wess-Zumino-Witten models. Indeed this precisely agrees with the $1$-form anomaly free conditions~\cite{Hung:2018rhg} of $3d$ Chern-Simons theories with general Lie groups, which justifies that the anomaly can flow from a bulk to a boundary. However the relation between invariant boundary state condition and mixed anomaly remains unclear. Can this $G$-invariant boundary state condition tell anything about the mixed anomaly between $G$ and other symmetries? In this paper we conjecture that if a $G$-invariant boundary state exists, not only the 't Hooft anomaly of $G$ is zero but also the mixed anomaly between $G$ and other symmetries is not allowed. We check this conjecture for known examples.

\section{Entanglement measure of mixed anomaly} \label{sec2}

\subsection{Entanglement form braiding}
Consider Chern-Simons theory with gauge group $\mathcal{G}$ at level $k$. The action is given by
\begin{equation}\label{csaction}
S_{CS}[A] = {k\over 4\pi} \int_M \Tr \left(AdA+{2\over 3}A^3\right)\ ,
\end{equation} where $M$ is a 3-manifold with a boundary $\partial M=\Sigma$. The Euclidean path integral of the theory on $M$ defines a quantum state on $\Sigma$. We choose $M$ as the link complement of a $n$-component link in $S^3$. There is a natural Hilbert space constructed as follows. One cuts along a tube neighborhood of the $n$-component link in $S^3$, then there are inside part and outside part, the solid torus and the link complement $M$. The inside path integral on a solid torus with a inserted Wilson loop (along the non-contractable circle) running over all integrable representations of affine algebra $\mathcal{G}_k$, provides a basis of a Hilbert space $\mathcal{H}$. Since there are $n$ disconnected inside parts, the total Hilbert space is $\mathcal{H}^{\otimes n}$. The state defined by the path integral on $M$ is
\begin{equation}
|\mathcal{L}\rangle \in \mathcal{H}^{\otimes n}\ .
\end{equation} We can assign a basis for the Hilbert space $\mathcal{H}$ so that the sate can be described specifically
\begin{equation}
|\mathcal{L}\rangle = \sum_{j_1,\cdots j_n} \omega(j_1,\cdots j_n)|j_1,\cdots j_n\rangle,
\end{equation} where $j$ runs over all the integrable representations and $|j\rangle$ is defined by the path integral on a solid torus with a Wilson loop in the representation $R^*_j$.~\footnote{Notice that this convention is different from that in~\cite{Balasubramanian:2016sro} up to a conjugation. This will not affect the later entropy counting.} Then the wave function $\omega(j_1,\cdots j_n)$ is
\begin{equation}
\omega(j_1,\cdots j_n)=\langle j_1,\cdots j_n|\mathcal{L}\rangle,
\end{equation} which is nothing but the expectation value of the Wilson loops in representation $R_{j_i}$ in $S^3$. This is because the inner product operationally means gluing in solid tori along the boundary of the link complement $S^3-\mathcal{L}$ and the fact that $\langle j_1,\cdots j_n|$ is the conjugate of $|j_1,\cdots j_n\rangle$. The equality between the wave function of the state on $\partial M=\Sigma$ and the VEV of linked Wilson loops allows us to assign a quantum entanglement structure to a link in $S^3$.

For our purpose, we only consider 2-component links from now on. In this case $M$ has a boundary of two tori. The wave function of the state defined on the two tori is given by the expectation value of two linked Wilson loops. In the case of Hopf link ($2$-component link with linking number $1$), the wave function is nothing but the modular $S$ matrix
\begin{equation}
|\mathcal{L}\rangle = \sum_{j_1, j_2} \mathcal{S}_{j_1,j_2}|j_1, j_2\rangle\ .
\end{equation}
 Notice that the Hilbert space constructed here is essentially the space of anyon species, which is very different from the Hilbert space in the computation of topological entanglement entropy~\cite{Levin:2006zz,Dong:2008ft,Kitaev:2005dm}. Given a modular $S$ matrix, unitarity requires
\begin{equation}
\rho = \mathcal{S}\mathcal{S}^\dagger = 1_{d\times d}\ ,
\end{equation} where $d$ is the dimension of a single torus Hilbert space. 
To build up a relation with the anomaly of the center $G$ of the affine algebra $\mathcal{G}_k$, one has to consider the truncated Hilbert space only consist of the $G$-symmetry lines (special anyons), then the wave function will be the truncated modular matrix $\hat{\mathcal{S}}$. The observation in~\cite{Hung:2018rhg} is that by tracing out one torus Hilbert space (truncated), the entanglement entropy measures precisely the order of the anomalous group
\begin{equation}\label{entropyanomaly}
\hat\rho = {\hat{\mathcal{S}}\hat{\mathcal{S}}^\dagger\over\Tr(\hat{\mathcal{S}}\hat{\mathcal{S}}^\dagger)}\ ,\quad S=-\Tr\hat\rho\log\hat\rho=\log D\ .
\end{equation} After truncation, $\hat{\mathcal{S}}$ is in general not unitary. If $\hat{\mathcal{S}}$ is unitary, we see that $D$ equals to the dimension of $\hat{\mathcal{S}}$ matrix therefore $G$ is fully anomalous. We have shown that 2 different conditions are equivalent
\begin{equation}
\text{Fully Anomalous}=~\text{Maximally Entangled}\ .
\end{equation} ``Maximally Entangled'' means the entanglement entropy reaches its maximal value.
One nontrivial point of (\ref{entropyanomaly}) is that the entanglement entropy still measures the anomaly even if $\hat{\mathcal{S}}$ is not unitary. Namely (\ref{entropyanomaly}) can measure the remaining anomalous symmetry even if $G$ is not fully anomalous.

\subsection{Mixed anomaly and residual entropy}
In this paper we focus on the mixed anomaly between $G_1$ and $G_2$. 
We propose an entropy measure of the mixed anomaly
\begin{equation}\label{proposal}
\Delta S = S[G_1\times G_2]-S[G_1]-S[G_2]\ .
\end{equation}
This measure can be viewed as the residual entropy between $G_1$ and $G_2$, namely the part not from individual $G_1$ or $G_2$ but from the mixing between the two.
The possible patterns of the anomaly for $G_1\times G_2$ can be classified as follows: 

 \begin{itemize}
\item 1)~~ $G_1$ and $G_2$ are both fully anomalous and there is no mixed anomaly between them. In this case following $S=\log D$ we have
\begin{equation}\label{opt1}
S[G_1]=\log d_1,\quad S[G_2]=\log d_2.
\end{equation} The order of the product group is $d[G_1\times G_2]=d_1 d_2$. Since the entanglement entropy measures the order of the anomalous group, it can not be greater than the order of the entire group,
\begin{equation}
S[G_1]+S[G_2]\leq S[G_1\times G_2]\leq \log (d_1d_2).
\end{equation}
Notice that the first inequality follows from that the total anomaly can not be less than the sum of individual ones. Together with (\ref{opt1}) we see that there is no residual entropy 
\begin{equation}
\Delta S := S[G_1\times G_2]-S[G_1]-S[G_2]=0\ .
\end{equation}
We interpret this result as that there can not exist mixed anomaly between $G_1$ and $G_2$.
\item 2)~~ $G_1$ and $G_2$ are both anomaly free but there is a mixed anomaly between them. In this case we have individually
\begin{equation}
S[G_1]=0,\quad S[G_2]=0\ ,
\end{equation}
but the mixed anomaly contributes to the 't Hooft anomaly of the product group $G_1\times G_2$
\begin{equation}
S[G_1\times G_2]>0\ .
\end{equation} So there is a finite residual entropy $\Delta S = S[G_1\times G_2]>0$.

\item 3)~~ Only one of the groups is anomaly free (say $G_1$) and there is a mixed anomaly between $G_1$ and $G_2$. In this case
\begin{equation}
S[G_1]=0,~S[G_2]>0, \quad S[G_1\times G_2]>S[G_2]\ .
\end{equation} The second inequality follows from the additivity of the anomaly. So there is a residual entropy $\Delta S = S[G_1\times G_2]-S[G_2]>0$. Notice that one can equivalently view this case as that one of the groups is anomalous and meanwhile there is a mixed anomaly between the two.

\item 4)~~One of the groups is anomaly free (say $G_1$) and also there is no mixed anomaly between the two. In this case, the total anomaly solely comes from $G_2$
\begin{equation}
S[G_1]=0\ ,\quad S[G_1\times G_2] = S[G_2]\ .
\end{equation} There is no residual entropy $\Delta S=0$. Whenever a subgroup is anomaly free and there is no mixed anomaly between this subgroup and other symmetry lines we call this subgroup anomaly decoupled. Anomaly decoupled subgroups will be irrelevant for the entropy counting. Notice that strictly
\begin{equation}
\text{Anomaly Free} \neq \text{Anomaly Decoupled}\ .
\end{equation} One can distinguish them through mixed anomaly. Anomaly free of a subgroup does not require no mixed anomaly with others but anomaly decoupled does.
\end{itemize}

Below we illustrate mixed anomalies in different theories following this classification.

\subsection{U(1)$_k$}
Abelian U(1) Chern-Simons theory at level $k$ has a global 1-form $Z_k$ symmetry~\cite{Gaiotto:2014kfa}. The symmetry lines are the Wilson lines
\begin{equation}
U_g(a)=\exp(i n\oint_a A)\ ,\quad g=e^{2\pi i n\over k}\ ,
\end{equation} with $n=0,1,\dots, k-1$. The charged operators are the same Wilson lines and the symmetry transformation rules are
\begin{equation}\label{U1act}
U_g(a)U_{g'}(b) = e^{2\pi i nm\over k} U_{g'}(b)\ ,\quad g=e^{2\pi i n\over k},~g'=e^{2\pi i m\over k}\ ,
\end{equation} where $a$ is a circle around $b$.
Equation (\ref{U1act}) means that the symmetry lines themselves are charged under $Z_k$. This has been interpreted as the 't Hooft anomaly of $Z_k$. Following the classification in the introduction, we choose the generator of $Z_k$ as
\begin{equation}
U=U_{e^{2\pi i /k}}=\exp(i \oint_a A) \quad\text{such that}\quad U^k=1\ ,
\end{equation} which is charged under itself
\begin{equation}
U(a)U(b) = e^{2\pi i \over k} U(b)\ .
\end{equation} This will set $q(U)=1$ in (\ref{symano}). One can in principle choose another line $U_g$ with $g=e^{2\pi i n\over k}$ and $\gcd(n,k)=1$. $U_g$ will also generate a $Z_k$ group and satisfy $U^k_g=1$. This means that the choice of $U$ is not unique in general for a given system. The value of $q(U)$ sometimes depends on the choice of the generator. Nevertheless one can use $q(U)$ to classify certain properties such as anomaly structures.

One can measure the expectation value of equation (\ref{U1act}) on a 3-manifold. For instance on $S^3$ the anomalous phase in (\ref{U1act}) is nothing but the modular S-matrix element (up to normalization). Therefore the anomaly can also be detected from the modular S matrix,
\begin{equation}\label{abelianS}
\mathcal{S}_{m,n}={1\over \sqrt{k}}e^{2\pi i mn\over k}\ .
\end{equation}
It has been observed~\cite{Hung:2018rhg} that the Von Neumann entropy of the reduced density matrix $\rho=\mathcal{S}\mathcal{S}^\dagger$ measures the order of anomalous group. Since the S matrix (\ref{abelianS}) is unitary, we have
\begin{equation}
S=-\tr\rho\log\rho=\log k\ ,
\end{equation} which reflects that $Z_k$ is fully anomalous. 

Now we check some mixed anomaly. 
\begin{itemize}
\item When $k=2(2\ell+1)$ ($\ell$ is integer), $Z_{2(2\ell+1)}=Z_2\times Z_{(2\ell+1)}$. The $Z_2$ symmetry is generated by $U_g$ with
\begin{equation}
g=e^{2\pi i (2\ell+1)\over k} = e^{\pi i} = -1\ .
\end{equation} The braiding phase of $U_g$ with itself is given by the $(2\ell+1,2\ell+1)$ element of the modular S matrix
\begin{equation}
e^{2\pi i (2\ell+1)^2\over k} = (-1)^{2\ell+1} = -1\ .
\end{equation} This will set $q(U)=1$ in (\ref{symano}) for $Z_2$.

Similarly the $Z_{(2\ell+1)}$ symmetry is generated by $U_{g'}$ with
\begin{equation}
g'=e^{2\pi i \times 2\over k} = e^{2\pi i\over 2\ell+1}\ .
\end{equation} The braiding phase of $U_{g'}$ and itself is given by the $(2,2)$ element of the modular S matrix
\begin{equation}
e^{2\pi i\times 2^2\over k} = e^{2\pi i\times2\over 2\ell+1}\ .
\end{equation} This will set $q(U)=2$ in (\ref{symano}) for $Z_{2\ell+1}$. In particular, $\gcd(2,2\ell+1)=1$ implies that $Z_{2\ell+1}$ is fully anomalous.

Now let us consider the mixed anomaly between $Z_2$ and $Z_{2\ell+1}$. The braiding phase between $U_g$ and $U_{g'}$ is
\begin{equation}
e^{2\pi i 2(2\ell+1)\over k} = 1\ .
\end{equation} This will set $\lambda=1$ in (\ref{lambda}). So there is no mixed anomaly between $Z_2$ and $Z_{2\ell+1}$. There is another quick way to see this through (\ref{mixc}) since there is no nontrivial solution for $M=2$ and $N=2\ell+1$ in that equation.
On the other hand, we know $Z_{2(2\ell+1)}$ is fully anomalous because of the unitarity of (\ref{abelianS}), so we have
\begin{equation}
S[Z_{2(2\ell+1)}]-S[Z_2]-S[Z_{(2\ell+1)}] = 0\ ,
\end{equation} where $S[G]$ is the entanglement entropy for the truncated S matrix consist of only $G$-symmetry lines. This provides a check of our proposal (\ref{proposal}).
\end{itemize}
Similar checks can be done for other subgroups of $Z_k$.

\subsection{SU$(N)_k$}
SU($N$) Chern-Simons theory at level $k$ has a global 1-form $Z_N$ symmetry~\cite{Gaiotto:2014kfa} associated to the center. The generator of this $Z_N $ symmetry is the Wilson line $U$ in a symmetric representation with $k$ boxes. Wilson lines in representations with $\ell$ boxes have charge $\ell$ under $U$. Therefore $U$ is charged under itself
\begin{equation}
U(a)U(b) = e^{2\pi i k\over N} U(b)\ .
\end{equation}
This will set $q(U)=k$ in (\ref{symano}) for $Z_N$. When $\gcd(N,k)=1$, this $Z_N$ symmetry is fully anomalous.

\subsubsection{$N=2(2\ell+1)$}
There are two subgroups $Z_2$ and $Z_{2\ell+1}$ and also $Z_{N}= Z_2\times Z_{2\ell+1}$. If we fix $k=1$, this is very similar to U$(1)_{2(2\ell+1)}$.
In general, the $Z_2$ generator is the Wilson line in representation with $(2\ell+1)k$ boxes. It is charged under itself with the braiding phase
\begin{equation}\label{Z2anomaly}
e^{{2\pi i k(2\ell+1)\over N}\times(2\ell+1)} = (-1)^k\ .
\end{equation} This will set $q(U)=k$ in (\ref{symano}) for $Z_2$.
The $Z_{2\ell+1}$ generator is the Wilson line in representation with $2k$ boxes. It is charged under itself with the braiding phase
\begin{equation}
e^{{2\pi i k\times 2\over N}\times2} = e^{{2\pi i}{2k\over 2\ell+1}}\ .
\end{equation} This will set $q(U)=2k$ in (\ref{symano}) for $Z_{2\ell+1}$.
For the mixed anomaly one has to consider the braiding of $Z_2$ generator and $Z_{2\ell+1}$ generator. The braiding phase is
\begin{equation}
e^{{2\pi ik\over N}2(2\ell+1)} = 1\ ,
\end{equation} so there is no mixed anomaly. Now let us check the residual entropy. When $\gcd(k,N)=1$, $Z_N$ is fully anomalous. In this case we will also have $\gcd(k,2)=1$ and $\gcd(k,2\ell+1)=1$ since $N=2(2\ell+1)$, so both $Z_2$ and $Z_{2\ell+1}$ are fully anomalous.
Following (\ref{entropyanomaly}) there is no residual entanglement entropy, namely $S[Z_{N}]-(S[Z_2]+S[Z_{2\ell+1}])=0$. This agrees with our proposal (\ref{proposal}).

Take SU$(6)_1$ as an example. The modular S matrix is given by
 \begin{equation}
\mathcal{S}={1\over \sqrt{6}}\left(
\begin{array}{cccccc}
 1 & 1 & 1 & 1 & 1 & 1 \\
 1 & \frac{1}{2} \left(1-i \sqrt{3}\right) & -\frac{1}{2} i \left(-i+\sqrt{3}\right) & -1 & \frac{1}{2} i \left(i+\sqrt{3}\right) & \frac{1}{2} \left(1+i \sqrt{3}\right) \\
 1 & -\frac{1}{2} i \left(-i+\sqrt{3}\right) & \frac{1}{2} i \left(i+\sqrt{3}\right) & 1 & -\frac{1}{2} i \left(-i+\sqrt{3}\right) & \frac{1}{2} i \left(i+\sqrt{3}\right) \\
 1 & -1 & 1 & -1 & 1 & -1 \\
 1 & \frac{1}{2} i \left(i+\sqrt{3}\right) & -\frac{1}{2} i \left(-i+\sqrt{3}\right) & 1 & \frac{1}{2} i \left(i+\sqrt{3}\right) & -\frac{1}{2} i \left(-i+\sqrt{3}\right) \\
 1 & \frac{1}{2} \left(1+i \sqrt{3}\right) & \frac{1}{2} i \left(i+\sqrt{3}\right) & -1 & -\frac{1}{2} i \left(-i+\sqrt{3}\right) & \frac{1}{2} \left(1-i \sqrt{3}\right) \\
\end{array}
\right)\ ,
\end{equation} where both $Z_2$ and $Z_3$ are anomalous but there is no mixed anomaly between them.

\subsubsection{$N=m\times n$}
In this case we have two subgroups $Z_n$ and $Z_m$. The generator of $Z_n$ is the Wilson line in the representation with $mk$ boxes. It is charged under itself by
\begin{equation}
U(a)U(b) = e^{2\pi i k m^2\over N} U(b)= e^{2\pi i k m\over n} U(b)\ ,
\end{equation} 
This will set $q(U)=km$. 
Similarly the $Z_m$ generator is charged under itself
\begin{equation}
U'(a)U'(b) = e^{2\pi i k n^2\over N} U'(b)= e^{2\pi i k n\over m} U'(b)\ .
\end{equation}
For the mixed anomaly we have
\begin{equation}\label{mixmn}
U(a)U'(b) = e^{2\pi i k mn\over N} U'(b)= U'(b)\ .
\end{equation}
Now consider the special case of $m=n$, $Z_m$ and $Z_n$ symmetry lines coincide. In this case (\ref{mixmn}) implies that the subgroup $Z_{n=m}$ is 't Hooft anomaly free. Let us consider the braiding of $Z_N$ generator  $U_0$ with $Z_n$ generator:
\begin{equation}\label{anomalyNn}
U^s_0(a)U(b) = e^{2\pi i k s n\over N} U(b) = e^{2\pi i k s\over n} U(b)\ . 
\end{equation} There is a mixed anomaly between $Z_n$ and the symmetry lines of $Z_N$ (except for $Z_n$ lines themselves). When $\gcd(k,N)=1$, $Z_N$ symmetry is fully anomalous.
On the other hand $Z_n$ is anomaly free. If there is no mixed anomaly between $Z_n$ and other symmetry lines, $Z_n$ will decouple and we will not have enough symmetry lines to match the total anomaly $\log N$. In another word, the finite residual entropy $\Delta S$ indicates that there must be a mixed anomaly, which is consistent with (\ref{anomalyNn}). One may call the mixed anomaly in this case the mixed anomaly between $Z_n$ and the quotient group $Z_N/Z_n$.

Take SU$(4)_1$ as an example. The modular S matrix is given by
 \begin{equation}
\mathcal{S}_{k=1}=\left(
\begin{array}{cccc}
 1 & 1 & 1 & 1 \\
 1 & -i & -1 & i \\
 1 & -1 & 1 & -1 \\
 1 & i & -1 & -i \\
\end{array}
\right)\ ,
\end{equation} where the subgroup $Z_2$ is anomaly free but there is a mixed anomaly between $Z_2$ and the quotient group.
\subsection{SO$(2N)_k$ with even $N$}
SO$(2N)$ Chern-Simons theory with even $N$ at level $k$ has a one-form symmetry $G=Z_2\times Z_2$. $G$ is anomaly free if and only if $k$ is even~\cite{Hung:2018rhg}. The truncated S matrix associated to symmetry lines is always a $4\times 4$ matrix. Therefore we turn to the truncated modular S matrix to discuss the relations between anomaly and entropy. Let us take $SO(8)_1$ as an example. In this case the modular S matrix is
\begin{equation}
\mathcal{S} = {1\over 2}\left(
\begin{array}{cccc}
 1 & 1 & 1 & 1 \\
 1 & 1 & -1 & -1 \\
 1 & -1 & 1 & -1 \\
 1 & -1 & -1 & 1 \\
\end{array}
\right)\ . 
\end{equation}
One can easily check that  both $Z_2$ subgroups are anomaly free since the corresponding $2\times 2$ matrix are identity matrices.~\footnote{One can choose any two lines among the three nontrivial lines as the generators of $Z_2\times Z_2$.} However, there is a nontrivial braiding phase $-1$ between two $Z_2$ symmetry lines. There is a mixed anomaly between them. Now let us check the entropy. The entanglement entropy for the total modular S matrix is
\begin{equation}
S=-\text{Tr}(\rho\log\rho)=\log 4, \quad \rho={\mathcal{S}\mathcal{S}^\dagger\over \text{Tr}\mathcal{S}\mathcal{S}^\dagger}.
\end{equation} Since the two $Z_2$ subgroups do not have entropy, $\Delta S=S[Z_2\times Z_2]$. The total finite entropy purely comes from the mixed anomaly between the two $Z_2$ subgroups.

Now let us turn to $SO(12)_1$. In this case, the modular S matrix is
\begin{equation}
\mathcal{S} = {1\over 2}\left(
\begin{array}{cccc}
 1 & 1 & 1 & 1 \\
 1 & -1 & 1 & -1 \\
 1 & 1 & -1 & -1 \\
 1 & -1 & -1 & 1 \\
\end{array}
\right)\ . 
\end{equation} The one-form symmetry is again $Z_2\times Z_2$. There are 4 symmetry lines. Apart from the identity, one can take any two among the three nontrivial lines as two $Z_2$ generators. Suppose we take the second and the third. Then we see from the modular S matrix, both of the two $Z_2$ are anomalous. And there is no mixed anomaly between them. If we instead take the second and the fourth lines, the former has anomaly but the latter does not have. Now there is a mixed anomaly between these two $Z_2$ symmetries. Let us now check the entropies. This is somewhat more intuitive. First of all, the total entropy out of this modular S matrix is $\log 4$, which is the same amount of the $SO(8)_1$ theory. Following (\ref{entropyanomaly}) the first choice of taking two anomalous $Z_2$ lines does not allow any residual entropy. Namely, $S[Z_2\times Z^\prime_2]-S[Z_2]-S[Z^\prime_2]=0$. This agrees with the fact that there is no mixed anomaly. In the second option, only one of the two lines is anomalous, therefore the residual entropy $S[Z_2\times Z^\prime_2]-S[Z_2]-S[Z^\prime_2]=\log 2$ is finite. This again agrees with the fact that there is a mixed anomaly now. In either case, the precise correspondence between the anomaly and the entanglement entropy holds. In particular, the residual entropy is indeed a good measure of the mixed anomaly.
\subsection{SO$(2N)_k$ with odd $N$}
SO$(2N)$ Chern-Simons theory with odd $N$ at level $k$ has a one-form symmetry $G=Z_4$. $G$ is anomaly free if and only if $k$ is $4\mathbb{Z}$~\cite{Hung:2018rhg}. The truncated S matrix associated to symmetry lines is $4\times 4$ matrix. Let us take SO$(10)_1$ as an example. The modular S matrix is
\begin{equation}
\mathcal{S} = {1\over 2}\left(
\begin{array}{cccc}
 1 & 1 & 1 & 1 \\
 1 & i & -i & -1 \\
 1 & -i & i & -1 \\
 1 & -1 & -1 & 1 \\
\end{array}
\right)\ .
\end{equation} From the modular S matrix, one can see there is a $Z_2$ subgroup of $Z_4$ generated by the fourth Wilson line. There are nontrivial phases from braiding this $Z_2$ line with the other lines, indicating there is mixed anomaly between $Z_2$ and the quotient group $Z_4/Z_2$.
\subsection{$Z_2$ gauge theory}
Another example is provided by the Kitaev toric code model, which can be described by the K-matrix Chern-Simons theory. The quasi-particle excitations (anyons) can be considered as the end of Wilson lines. If a quasi-particle is dragged around another quasi-particle, the Wilson line attached to them may link with each other. The braiding phase between Wilson lines is essentially the braiding phase between quasi-particles. The fundamental excitations in Kitaev toric code model are $Z_2$ charge $e$ and flux $m$. They are bosons but can combine to form a fermionic composite quasi-particle $\psi=e\times m$. The S and T matrices are 
\begin{equation}
\mathcal{S} = {1\over 2}\left(
\begin{array}{cccc}
 1 & 1 & 1 & 1 \\
 1 & 1 & -1 & -1 \\
 1 & -1 & 1 & -1 \\
 1 & -1 & -1 & 1 \\
\end{array}
\right)\ ,\quad
\mathcal{T}= \text{diag} (1,1,1,-1).
\end{equation} This modular S matrix is the same as that of SO$(8)_1$ theory. There is a $Z_2\times Z_2$ one-form symmetry acted by braiding quasi-particles. There is a mixed anomaly between the two $Z_2$ though both of them are anomaly free. Since the edge theory is explicitly known, one can also study the mixed anomaly in the $2d$ edge theory. 

Similar mixed anomaly also exists in $3d$ $Z_N$ gauge theory. The theory has a $Z_N\times Z_N$ one-form symmetry, generated by the basic electric and magnetic lines. Both of electric and magnetic $Z_N$ are anomaly free but there is a mixed anomaly between them. This can be seen from that electric and magnetic lines have a mutual braiding phase $e^{2\pi i\over N}$.

\section{Boundary states and mixed anomaly}
In this section we discuss the relation between mixed anomaly and invariant boundary state condition. The $2d$ counterpart of $3d$ Chern-Simons theory is Wess-Zumino-Witten models. The $1$-form symmetry of Chern-Simons theory becomes $0$-form symmetry in Wess-Zumino-Witten models, because the $1$-dimensional symmetry lines in $3d$ becomes co-dimension $1$ topological defect lines in $2d$. Therefore the $1$-form anomaly of Chern-Simons theory becomes $0$-form anomaly in Wess-Zumino-Witten models. The 't Hooft anomaly of a $0$-form $G$-symmetry in bosonic $2d$ theory is classified by the cohomology group $H^3(G,U(1))$. In~\cite{Han:2017hdv}, it has been proposed that one can use $G$-invariant boundary state condition to detect the 't Hooft anomaly, namely the existence of a $G$-invariant boundary state will imply that the theory is $G$-anomaly free. By this mean,~\cite{Numasawa:2017crf} computed the anomaly free condition (conditions on the level $k$) for the center symmetry in Wess-Zumino-Witten models, which precisely agrees with the $1$-form anomaly in Chern-Simons theory with general affine algebra~\cite{Hung:2018rhg}. We will focus on the relation between invariant boundary state condition and mixed anomaly. 

We first review the invariant boundary state condition in WZW models~\cite{Han:2017hdv,Numasawa:2017crf} for later discussions. 
To form the physical boundary states of Wess-Zumino-Witten models, we need a basis called Ishibashi states. Since the Ishibashi states are certain linear combinations of the primary states $|\hat\lambda,\hat\lambda\rangle$ and their descendants, we can formally denote them as
\begin{equation}
|\hat\lambda\rangle\rangle\ ,\quad \hat\lambda\in P^k_+\ ,
\end{equation}  where $\hat\lambda$ is the level $k$ integrable highest weights $\hat\lambda = (\lambda_1,\cdots,\lambda_N)$ of the affine albegra and the finite set $P^k_+$ is defined in terms of the comarks $a_i$ by
\begin{equation}
P^k_+ = \{ (\lambda_1,\cdots,\lambda_N) | a_1\lambda_1+\cdots+a_N\lambda_N\leq k,~\lambda_i\geq 0\}\ .
\end{equation}
The physically realized Cardy states are given by
\begin{equation}
|\hat\mu\rangle_c=\sum_{\hat\lambda\in P^k_+}{S_{\hat\mu,\hat\lambda}\over \sqrt{S_{\hat 0,\hat\lambda}}}|\hat\lambda\rangle\rangle\ ,\quad \hat\mu\in P^k_+\ ,
\end{equation}where $S_{\hat\mu,\hat\lambda}$ is the modular S matrix. In order to consider the $G$-invariant boundary state condition, we have to specify the $G$-symmetry action on the Cardy states. The transformation is through the Ishibashi states. Let us denote the generator of $G$ by $g$, then
\begin{equation}
g|\hat\mu\rangle_c = \sum_{\hat\lambda\in P^k_+}{S_{\hat\mu,\hat\lambda}\over \sqrt{S_{\hat 0,\hat\lambda}}}\,g|\hat\lambda\rangle\rangle\ .
\end{equation} Recall that the action of $g$ on the primaries are given by
\begin{equation}
g|\hat\lambda,\hat\lambda\rangle = e^{-2\pi i (A\hat\omega_0,\lambda)}|\hat\lambda,\hat\lambda\rangle\ ,
\end{equation} and the outer-automorphism acting on the modular S matrix is given by~\footnote{For the derivation we refer to Page 595 of ``Conformal Field Theory'' by Francesco, Mathieu and Senechal. We use the same notation.}
\begin{equation}
AS_{\hat\mu,\hat\lambda} := S_{A\hat\mu,\hat\lambda} = S_{\hat\mu,\hat\lambda}e^{-2\pi i (A\hat\omega_0,\lambda)}\ .
\end{equation} The center symmetry rotation on Cardy states is then given by the outer automorphism,
\begin{equation}
g|\hat\mu\rangle_c = |A\hat\mu\rangle_c\ .
\end{equation} Therefore to find a state invariant under $g$ is  equivalent to find a affine weight satisfying 
\begin{equation}
|A\hat\mu\rangle_c = |\hat\mu\rangle_c\ .
\end{equation}

\subsection{SU$(2N)_k$}
Consider SU$(2N)_k$ WZW models. The affine Dynkin labels are $[\lambda_0; \lambda_1,\dots,\lambda_{2N-2},\lambda_{2N-1}]$.
\begin{equation}
Z_{2N}~\text{rotation}: A[\lambda_0; \lambda_1,\cdots,\lambda_{2N-2},\lambda_{2N-1}] =  [\lambda_{2N-1}; \lambda_0,\cdots,\lambda_{2N-3},\lambda_{2N-2}]\ .
\end{equation}
Equating the Dynkin labels before and after the rotation, one obtains
\begin{equation}
\lambda_0=\lambda_1=\cdots=\lambda_{2N-2}=\lambda_{2N-1}\ ,
\end{equation} which constrains the level
\begin{equation}
k:=\lambda_0+\sum_{j=1}^{2N-1}\lambda_j = 2N\lambda_0\in 2N\mathbb{Z}\ .
\end{equation}
Now let us consider a rotation of $Z_2$ subgroup
\begin{equation}
Z_{2}~\text{rotation}: A[\lambda_0; \lambda_1,\cdots,\lambda_N,\cdots,\lambda_{2N-2},\lambda_{2N-1}] =  [\lambda_{N}; \lambda_{N+1},\cdots,\lambda_0,\cdots,\lambda_{N-2},\lambda_{N-1}]\ .
\end{equation}
Equating the Dynkin labels before and after the $Z_2$ rotation, one obtains
\begin{equation}
\lambda_0=\lambda_N; \lambda_1=\lambda_{N+1};\cdots;\lambda_{N-1}=\lambda_{2N-1};
\end{equation} which constrains the level
\begin{equation}
k:=\lambda_0+\sum_{j=1}^{2N-1}\lambda_j = 2(\lambda_0+\lambda_1+\cdots+\lambda_{N-1}) \in 2\mathbb{Z}\ .
\end{equation}
One can also analyze the anomaly condition using (\ref{symano}) introduced in the introduction. The braiding phase of the $Z_2$ line with itself is
\begin{equation}\label{Z2Nk}
U(a)U(b) = e^{2\pi i N^2k\over 2N}U(b)=(-1)^{Nk}U(b)\ .
\end{equation} $k\in2\mathbb{Z}$ is a sufficient but not necessary condition for $Z_2$ anomaly free. Now consider the $Z_2$ line braiding with the $Z_{2N}$ generator $U_0$,
\begin{equation}\label{Z2k}
U(a)U_0(b) = e^{2\pi i Nk\over 2N}U_0(b) = (-1)^kU_0(b)\ ,
\end{equation} from which we see that $Z_2$ line actually becomes decoupled when $k$ is even.
Compare (\ref{Z2Nk}) and (\ref{Z2k}), we find that $k\in2\mathbb{Z}$ not only makes $Z_2$ anomaly free but also makes it anomaly decoupled. In another word, the $Z_2$-invariant boundary state condition will guarantee $Z_2$ anomaly decoupled.
Actually for any subgroup $Z_n$, one can check that the invariant boundary state condition will constrain $k\in n\mathbb{Z}$. This is precisely the condition that makes the subgroup $Z_n$ decoupled. Namely there can not be mixed anomaly between $Z_n$ and other symmetries. From the above analysis, we conjecture that {\it $G$-invariant boundary state condition does not allow mixed anomaly between $G$ and other symmetries (internal gobal $0$-form).}

\subsection{SO$(2N)_k$ with even $N$}
Consider SO$(2N)_k$ WZW models. The one-form symmetry is $Z_2\times \widetilde Z_2$. The affine Dynkin labels are $[\lambda_0; \lambda_1,\dots,\lambda_{N-1},\lambda_{N}]$. Consider first the $Z_2$ rotation,
\begin{equation}
Z_{2}~\text{rotation}: A[\lambda_0; \lambda_1,\lambda_2,\cdots,\lambda_{N-2},\lambda_{N-1},\lambda_{N}] =  [\lambda_1; \lambda_0,\lambda_2,\cdots,\lambda_{N-2},\lambda_N,\lambda_{N-1}]\ .
\end{equation}
Equating the Dynkin labels before and after the rotation, one obtains
\begin{equation}
\lambda_0=\lambda_1; \lambda_{N-1}=\lambda_N;
\end{equation}
which constrains the level
\begin{equation}\label{SO2k}
k:=\lambda_0+\lambda_1+2\sum_{j=2}^{N-2}\lambda_j+\lambda_{N-1}+\lambda_{N} = 2(\lambda_0+\lambda_2+\cdots+\lambda_{N-2}+\lambda_{N-1}) \in 2\mathbb{Z}\ .
\end{equation}
Now consider the $\widetilde Z_2$ rotation,
\begin{equation}
\widetilde Z_{2}~\text{rotation}: A[\lambda_0; \lambda_1,\lambda_2,\cdots,\lambda_{N-2},\lambda_{N-1},\lambda_{N}] =  [\lambda_N; \lambda_{N-1},\lambda_{N-2},\cdots,\lambda_2,\lambda_1,\lambda_0]\ .
\end{equation}
Equating the Dynkin labels before and after the rotation, one obtains
\begin{equation}
\lambda_0=\lambda_N; \lambda_1=\lambda_{N-1};\cdots;\lambda_{{N\over 2}-1}=\lambda_{{N\over 2}+1};
\end{equation}
which constrains the level
\begin{equation}\label{SO2k1}
k:=\lambda_0+\lambda_1+2\sum_{j=2}^{N-2}\lambda_j+\lambda_{N-1}+\lambda_{N} = 2(\lambda_0+\lambda_1+2\sum_{j=2}^{{N\over 2}-1}\lambda_j+\lambda_{N\over 2}) \in 2\mathbb{Z}\ .
\end{equation}
Notice that either (\ref{SO2k1}) or (\ref{SO2k}) is enough to reproduce the anomaly free condition for the full group $Z_2\times \widetilde Z_2$. This agrees with our conjecture that, $G$-invariant boundary state condition not only makes $G$ anomaly free but also does not allow mixed anomaly between $G$ and other symmetries.

\section{Discussion}
't Hooft anomaly is important because it is preserved along renormalization group flows. How to detect the 't Hooft anomaly is an interesting question both for theorists and experimentalists. In~\cite{Hung:2018rhg} a concrete relation between 't Hooft anomaly and entanglement entropy has been proposed. This enables us to detect the anomaly of symmetry $G$ by measuring the entanglement entropy $S[G]$ for a state on a linked two tori. In this paper we generalize the idea in~\cite{Hung:2018rhg} to the mixed anomaly for $3d$ one-form symmetries, namely the residual entropy $\Delta S:=S[G_1\times G_2]-S[G_1]-S[G_2]$ can measure the mixed anomaly. This new relation shows that mixed anomaly can also be stored in the entropy, which provides further evidence for the correspondence between entanglement and anomaly in topological field theories. 

Although entanglement and anomaly are both purely quantum effects, they are generally very different. We often define anomaly using operator equations but define entanglement using states. Our concrete result between anomaly and entanglement entropy suggests that in order to find the quantum information counter part of usual observables in quantum field theories one has to switch to Hilbert space supported by conserved quantum numbers. In our case the truncated Hilbert space consist of all the symmetry lines is finite dimensional, and it is obviously interesting to generalize the relation between entropy and anomaly to infinite dimensional Hilbert space. This will allow us to find the analogous relation for the continuous symmetry such as $U(1)$.

Many $3d$ topological field theories have corresponding $2d$ conformal field theories. The $2d$ counterpart of $3d$ $1$-form symmetry is $0$-form. $2d$ global symmetries by themselves are very rich and how to detect the anomaly is also quite interesting~\cite{Chang:2018iay,Bhardwaj:2017xup}. For bosonic $2d$ theories the 't Hooft anomaly is classified by $H^3(G,U(1))$. Our $3d$ analysis brings insights to find general $2d$ criterions to detect both 't Hooft anomaly and also the mixed anomaly between two groups~\cite{KZ2019}. 

\acknowledgments
We would like to thank Ling Yan Hung, Ho Tat Lam, Liang Kong, Yifan Wang and Yong Shi Wu for useful discussions and Ken Kikuchi for related collaboration.

\appendix
\section{Level dependence}
In this appendix we illustrate the level $k$ dependence of the 't Hooft anomalies in Chern-Simons theories with general Lie groups.
\subsection{SU$(N)_k$}
In SU$(N)_k$ Chern-Simons theory, the $1$-form symmetry is the center $Z_N$. The braiding of the generator with itself
\begin{equation}
U(a)U(b)=e^{2\pi i k\over N}U(b)\ .
\end{equation} When $k$ and $N$ have common divisor, the faithful symmetry among the symmetry lines becomes $Z_{N/\gcd(k,N)}$. This implies the entanglement entropy for this $Z_N$ is
\begin{equation}\label{levelS}
S=\log{N\over \gcd(k,N)}\ ,
\end{equation} where the one-form symmetry is anomaly free only when $k$ is a multiplier of $N$. Let us illustrate a few examples to see that (\ref{levelS}) is correct. Take SU$(4)_k$ as examples. The truncated modular S matrices are 
\begin{equation}
\mathcal{S}_{k=1}=\left(
\begin{array}{cccc}
 1 & 1 & 1 & 1 \\
 1 & -i & -1 & i \\
 1 & -1 & 1 & -1 \\
 1 & i & -1 & -i \\
\end{array}
\right)\ ,\quad \mathcal{S}_{k=2}={1\over 2\sqrt{6}}\left(
\begin{array}{cccc}
 1 & 1 & 1 & 1 \\
 1 & -1 & 1 & -1 \\
 1 & 1 & 1 & 1 \\
 1 & -1 & 1 & -1 \\
\end{array}
\right)\ .
\end{equation}
\begin{equation}
\mathcal{S}_{k=3}=c\left(
\begin{array}{cccc}
 1 & 1 & 1 & 1 \\
 1 & i & -1 & -i \\
 1 & -1 & 1 & -1 \\
 1 & -i & -1 & i \\
\end{array}
\right)\ ,\quad \mathcal{S}_{k=4}={2-\sqrt{2}\over 16}\left(
\begin{array}{cccc}
 1 & 1 & 1 & 1 \\
 1 & 1 & 1 & 1 \\
 1 & 1 & 1 & 1 \\
 1 & 1 & 1 & 1 \\
\end{array}
\right)\ . 
\end{equation} One can check from the modular S matrices that $S=\log{N\over \gcd(k,N)}$ is true. Let us look at $SU(4)_2$. In the truncated $4\times4$ modular S matrix, the third row(column) is the symmetry line generating the $Z_2$ subgroup of $Z_4$, which is anomaly free. Further more, there is no nontrivial phase from the braiding of this line and others therefore there is no mixed anomaly between this $Z_2$ and others. We call $Z_2$ decoupled. In this case, only the quotient group $Z_4/Z_2$ is anomalous and this agrees with the entanglement entropy $S=\log{4\over \gcd(4,k=2)}=\log 2$. We emphasize that whenever a subgroup is anomaly free and does not contribute any mixed anomaly, then it is decoupled from the anomalous group. 

Take SU$(6)_k$ as examples. The modular S matrices are
\begin{equation}
\mathcal{S}_{k=1}={1\over\sqrt{6}}\left(
\begin{array}{cccccc}
 1 & 1 & 1 & 1 & 1 & 1 \\
 1 & \frac{1}{2} \left(1-i \sqrt{3}\right) & -\frac{1}{2} i \left(-i+\sqrt{3}\right) & -1 & \frac{1}{2} i \left(i+\sqrt{3}\right) & \frac{1}{2} \left(1+i \sqrt{3}\right) \\
 1 & -\frac{1}{2} i \left(-i+\sqrt{3}\right) & \frac{1}{2} i \left(i+\sqrt{3}\right) & 1 & -\frac{1}{2} i \left(-i+\sqrt{3}\right) & \frac{1}{2} i \left(i+\sqrt{3}\right) \\
 1 & -1 & 1 & -1 & 1 & -1 \\
 1 & \frac{1}{2} i \left(i+\sqrt{3}\right) & -\frac{1}{2} i \left(-i+\sqrt{3}\right) & 1 & \frac{1}{2} i \left(i+\sqrt{3}\right) & -\frac{1}{2} i \left(-i+\sqrt{3}\right) \\
 1 & \frac{1}{2} \left(1+i \sqrt{3}\right) & \frac{1}{2} i \left(i+\sqrt{3}\right) & -1 & -\frac{1}{2} i \left(-i+\sqrt{3}\right) & \frac{1}{2} \left(1-i \sqrt{3}\right) \\
\end{array}
\right)\ ,
\end{equation}
\begin{equation}
\mathcal{S}_{k=2}=c*\left(
\begin{array}{cccccc}
 1 & 1 & 1 & 1 & 1 & 1 \\
 1 & -\frac{1}{2} i \left(-i+\sqrt{3}\right) & \frac{1}{2} i \left(i+\sqrt{3}\right) & 1 & -\frac{1}{2} i \left(-i+\sqrt{3}\right) & \frac{1}{2} i \left(i+\sqrt{3}\right) \\
 1 & \frac{1}{2} i \left(i+\sqrt{3}\right) & -\frac{1}{2} i \left(-i+\sqrt{3}\right) & 1 & \frac{1}{2} i \left(i+\sqrt{3}\right) & -\frac{1}{2} i \left(-i+\sqrt{3}\right) \\
 1 & 1 & 1 & 1 & 1 & 1 \\
 1 & -\frac{1}{2} i \left(-i+\sqrt{3}\right) & \frac{1}{2} i \left(i+\sqrt{3}\right) & 1 & -\frac{1}{2} i \left(-i+\sqrt{3}\right) & \frac{1}{2} i \left(i+\sqrt{3}\right) \\
 1 & \frac{1}{2} i \left(i+\sqrt{3}\right) & -\frac{1}{2} i \left(-i+\sqrt{3}\right) & 1 & \frac{1}{2} i \left(i+\sqrt{3}\right) & -\frac{1}{2} i \left(-i+\sqrt{3}\right) \\
\end{array}
\right)\ ,
\end{equation}
\begin{equation}
\mathcal{S}_{k=3}=\tilde c*\left(
\begin{array}{cccccc}
 1 & 1 & 1 & 1 & 1 & 1 \\
 1 & -1 & 1 & -1 & 1 & -1 \\
 1 & 1 & 1 & 1 & 1 & 1 \\
 1 & -1 & 1 & -1 & 1 & -1 \\
 1 & 1 & 1 & 1 & 1 & 1 \\
 1 & -1 & 1 & -1 & 1 & -1 \\
\end{array}
\right)\ .
\end{equation} One can easily check that, the entanglement entropy at different levels $k=1,2,3$ are $S=\log 6\ ,\log 3\ ,\log2$, respectively, which agrees with the general formula $S=\log{N\over \gcd(N,k)}$. When $k=2$, the subgroup $Z_2$ is decoupled and when $k=3$, the subgroup $Z_3$ is decoupled.
\subsection{Other types}
For SO$(2N+1)_k$ theories, the 1-form symmetry is $Z_2$. This $Z_2$ is always anomaly free so there is no $k$ dependence. For Sp$(2N)_k$ theories the one-form symmetry is $Z_2$ and it is anomalous when $Nk$ is odd. Therefore the entropy can be written as $S=\log{2\over \gcd(2,Nk)}$. For SO$(2N)_k$ theories, the one-form symmetry is $Z_2\times Z_2$ when $N$ is even and $Z_4$ when $N$ is odd. The anomaly free condition is $k\in 2\mathbb{Z}$ when $N$ is even and $k\in 4\mathbb{Z}$ when $N$ is odd. When $N$ is even, $k\in 2\mathbb{Z}$ is enough to make both $Z_2$ groups anomaly free and also no mixed anomaly. For odd $N$, $S=\log{4\over \gcd(4,k)}$. Let us take SO$(10)_k$ as examples. The modular S matrices are
\begin{equation}
\mathcal{S}_{k=1} = {1\over 2}\left(
\begin{array}{cccc}
 1 & 1 & 1 & 1 \\
 1 & i & -i & -1 \\
 1 & -i & i & -1 \\
 1 & -1 & -1 & 1 \\
\end{array}
\right)\ ,\quad \mathcal{S}_{k=2}={1\over 2\sqrt{10}}\left(
\begin{array}{cccc}
 1 & 1 & 1 & 1 \\
 1 & -1 & -1 & 1 \\
 1 & -1 & -1 & 1 \\
 1 & 1 & 1 & 1 \\
\end{array}
\right)\ .
\end{equation}
\begin{equation}
\mathcal{S}_{k=3}=c\times\left(
\begin{array}{cccc}
 1 & 1 & 1 & 1 \\
 1 & -i & i & -1 \\
 1 & i & -i & -1 \\
 1 & -1 & -1 & 1 \\
\end{array}
\right)\ ,\quad
\mathcal{S}_{k=4}={2-\sqrt{3}\over 24}\left(
\begin{array}{cccc}
 1 & 1 & 1 & 1 \\
 1 & 1 & 1 & 1 \\
 1 & 1 & 1 & 1 \\
 1 & 1 & 1 & 1 \\
\end{array}
\right)\ .
\end{equation} One can check from these modular S matrices, the $k$ dependence of the entanglement entropy is given by
\begin{equation}
S=\log{4\over \gcd(4,k)}\ .
\end{equation}

\section{Linking number dependence}
In this appendix we summarize the linking number dependence of the entanglement entropy.
In previous definition of 1-form symmetry of $3d$ Chern-Simons theory, the symmetry line $U$ acts on the charged line $V$ by linking it once. However this is not a unique definition of a symmetry transformation. One can also define a symmetry transformation of $V$ by braiding $n$ times (with linking number $n$). Below we compare this newly defined symmetry and the original symmetry (with linking number $1$).
\subsection{U$(1)_k$}
For U$(1)$ Chern-Simons theory with level $k$, the 1-form symmetry is $Z_k$ if the symmetry is defined by linking once. The symmetry lines stay in representations $q=0,1,\cdots,k-1$. The modular S matrix can be identified as the wave function of a state on two linked torus boundaries,
\begin{equation}
\mathcal{S}_{q_1,q_2}={1\over \sqrt{k}}e^{2\pi i q_1q_2\over k}\ ,
\end{equation} where $q_1$ and $q_2$ span integers $0\leq q<k$. 
Recall that the modular T matrix in this case is given by
\begin{equation}
\mathcal{T}_{q_1,q_2} = e^{2\pi i h_{q_1}}\delta_{q_1,q_2}\ ,
\end{equation} where $h_q={q^2\over 2k}$. Now we consider a 2-component link with linking number $\ell$.
The wave function is given by~\cite{Witten:1988hf}
\begin{equation}\label{Slink}
\widetilde{\mathcal{S}}_{q_1,q_2}={1\over \sqrt{k}}e^{{2\pi i q_1q_2\over k}\ell}\ .
\end{equation} The entanglement entropy for the wave function with linking number $\ell$ can be computed from the reduced density matrix
\begin{equation}
\rho = {\widetilde{\mathcal{S}}\widetilde{\mathcal{S}}^\dagger \over \text{Tr} \widetilde{\mathcal{S}}\widetilde{\mathcal{S}}^\dagger}\ .
\end{equation} The Von Neumann entropy of $\rho$ is given by~\cite{Balasubramanian:2016sro}
\begin{equation}\label{entropylink}
S=-\text{Tr} \rho\log\rho=\ln \left(k\over \gcd(k,\ell)\right)\ .
\end{equation}
This reminds us the anomaly condition of $Z_k$ symmetry. Let us choose the symmetry line with charge $q=1$ as the $Z_k$ generator. The generator has braiding with itself
\begin{equation}
U(a)U(b) = e^{2\pi i \ell}U(b)\ .
\end{equation} This will set $q(U)=\ell$ in (\ref{symano}). Following our previous analysis, a subgroup $Z_{\gcd(k,\ell)}$ will be decoupled, which explains the entropy result.
(\ref{entropylink}) can be understood in a more intuitive way.
From the wave function (\ref{Slink}) one can see that $\ell$-linking is equivalent to $\ell$ times of linking by the same symmetry line (with linking number 1). This is because the symmetry is abelian.
$\ell$ times of symmetry transformations will of course decouple $Z_{\gcd(k,\ell)}$. The remaining faithful symmetry has order ${k\over \gcd(k,\ell)}$ as detected by the entropy (\ref{entropylink}).

\subsection{SU(2)$_k$}
For $SU(2)$ Chern-Simons theory with level $k$, the 1-form symmetry is $Z_2$ with symmetry lines staying in the representations with spin $0$ and spin ${k\over 2}$. The modular S matrix is given by,
\begin{equation}
\mathcal{S}_{j_1,j_2}=\sqrt{2\over k+2} \sin\left(\pi(2j_1+1)(2j_2+1)\over k+2\right)\ ,
\end{equation} where $j_1$ and $j_2$ span $0,{1\over 2},\dots,{k\over2}$. Now we only consider a $2\times 2$ truncated matrix with indices $j_{1,2}=0,{k\over 2}$. We denote the new matrix as $\hat{\mathcal{S}}$,
\begin{equation}
\hat{\mathcal{S}} = {\sqrt{2\over k+2}}\sin\left(\pi\over k+2\right)\left(\begin{array}{cc}
1 & 1 \\
 1 & {\sin\left(k\pi + {\pi\over k+2}\right)\over \sin\left(\pi\over k+2\right)} \\
\end{array} \right)\ .
\end{equation}
Recall the modular T matrix in this case is given by
\begin{equation}
\mathcal{T}_{j_1,j_2} = e^{2\pi i h_{j_1}}\delta_{j_1,j_2}\ ,
\end{equation} where $h_j={j(j+1)\over k+2}$. After truncation it is given by
\begin{equation}
\hat{\mathcal{T}} = \left(
\begin{array}{cc}
1 & 0 \\
 0 & e^{i{\pi k\over 2}} \\
\end{array}
\right)\ .
\end{equation}
The wave function of a 2-component link with linking number $\ell$ in non-abelian SU$(2)_k$ theory can be worked out from the surgery method~\cite{Witten:1988hf}
\begin{equation}
\mathcal{P}_{j_1,j_2} = \sum_k (\mathcal{S}\mathcal{T}^\ell \mathcal{S})_{0k}{\mathcal{S}_{j_1k}\mathcal{S}_{j_2k}\over \mathcal{S}_{0k}}\ .
\end{equation} For our purpose we instead use the truncated matrices $\hat{\mathcal{S}}$ and $\hat{\mathcal{T}}$ because we are only interested in the anomaly structure associated to the $Z_2$ symmetry. This gives us a wave function $\hat{\mathcal{P}}$:
\begin{equation}
\hat{\mathcal{P}}_{j_1,j_2}\sim\left(
\begin{array}{cc}
 \left(1+(-1)^k\right) e^{\frac{i \ell k \pi }{2}}+2 & \left(1+(-1)^{2 k}\right) e^{\frac{i \ell k \pi }{2}}+(-1)^k+1 \\
 \left(1+(-1)^{2 k}\right) e^{\frac{i \ell k \pi }{2}}+(-1)^k+1 & \left(1+(-1)^{3 k}\right)e^{\frac{i \ell k \pi }{2}}+(-1)^{2 k}+1 \\
\end{array}
\right)\ .
\end{equation} When $k$ is even, there is no $Z_2$ anomaly and $\hat{\mathcal{P}}$ does not give finite entropy. When $k$ is odd, 
\begin{equation}
\hat{\mathcal{P}}_{j_1,j_2}\sim\left(
\begin{array}{cc}
2 & 2 e^{\frac{i \ell k \pi }{2}} \\
2 e^{\frac{i \ell k \pi }{2}} & 2 \\
\end{array}
\right)\ .
\end{equation}
The entanglement entropy of $\hat{\mathcal{P}}$ can be computed from the reduced density matrix~\footnote{It is normalization independent.}
\begin{equation}
\rho = {\hat{\mathcal{P}} \hat{\mathcal{P}}^\dagger \over \text{Tr} \hat{\mathcal{P}} \hat{\mathcal{P}}^\dagger}=\left(
\begin{array}{cc}
 \frac{1}{2} & \frac{1}{2} \cos ({k \ell \pi\over 2} ) \\
 \frac{1}{2} \cos ({k \ell \pi\over 2} ) & \frac{1}{2} \\
\end{array}
\right)\ .
\end{equation} The Von Neumann entropy of $\rho$ is given by
\begin{equation}
S=-\text{Tr} \rho\log\rho=\log{2\over \gcd(2,\ell)}\ ,\quad k~\text{odd}.
\end{equation}
$\ell=1$ is the usual case, where the entropy vanishes for even $k$ and equals to $\log 2$ for odd $k$. This agrees with the $k$ dependence (\ref{levelS}).
When $\ell\in 2\mathbb{Z}$, the entropy becomes zero.

\subsection{SU$(N)_k$}
For SU$(N)$ Chern-Simons theory with level $k$, the 1-form symmetry is $Z_N$ with symmetry lines staying in the representations with rectangular Young tableaux with $nk$ boxes. One can use a set of non-negative integer, the so called Dynkin labels $(a_1,\cdots, a_{N-1})$ to label each integrable representation $R$. The integrable representations are constrained by $\phi_1 a_1+\cdots+\phi_{N-1}a_{N-1}\leq k$, where $k$ is the Chern-Simons level and $(\phi_1,\cdots, \phi_{N-1})$ are the comarks. For su$(N)$ algebra this is given by $(1,\cdots,1)$. The modular S matrix is a matrix in the Hilbert space including all integrable representations. For two representations, $a=(a_1,\cdots, a_{N-1})$ and $b=(b_1,\cdots, b_{N-1})$, the modular S matrix is given by
\begin{equation}
\mathcal{S}_{a,b}=(-i)^{N(N-1)\over 2}{N^{-1/2}\over (N+k)^{N-1\over 2}}\text{Det}_{ij}\left[\exp\left({2\pi i \phi_{a}[i]\phi_{b}[j]\over N+k}\right)\right]\ .
\end{equation} where $i,j =1,\cdots, N$. For a given representation, the function $\phi$ is
\begin{equation}
\phi_a[i] = L_i-i-{L\over N}+{N+1\over 2}\ ,
\end{equation} where $L_i=\sum_{m=i}^{N-1} a_m$ and $L=\sum_{m=1}^{N-1} L_m$. The modular T matrix is given by
\begin{equation}
\mathcal{T}_{a,b}=\delta_{a,b}\exp\left(-2\pi i {k(N^2-1)\over 24 (N+k)}\right)\exp\left[{2\pi i\over 2(N+k)}\left({x_a(N^2-x_a)\over N}+y_a\right)\right]\ ,
\end{equation} where
$x_a$ and $y_a$ are integers determined by Dynkin labels
\begin{equation}
x_a = \sum_{i=1}^{N-1} i a_i\ ;\quad  y_a =\sum_{i=1}^{N-1}a_i\left(-i^2+\sum_{j=1}^i j a_j + \sum_{j=i+1}^{N-1} i a_j\right)\ .
\end{equation}
Truncation means we only consider a $N\times N$ matrix with representations labeled by rectangular Young tableaux with $L=nk$ boxes. We denote the truncated S matrix as $\hat{\mathcal{S}}$. Similarly the truncated T matrix is a $N\times N$ matrix with row and column corresponding to those of $\hat{\mathcal{S}}$. We denote it as $\hat{\mathcal{T}}$.
The wave function of a two component link with linking number $\ell$ is given by
\begin{equation}
\mathcal{P}_{j_1,j_2} = \sum_k (\mathcal{S}\mathcal{T}^\ell \mathcal{S})_{0k}{\mathcal{S}_{j_1k}\mathcal{S}_{j_2k}\over \mathcal{S}_{0k}}\ .
\end{equation} For our purpose we instead use $\hat{\mathcal{S}}$ and $\hat{\mathcal{T}}$.
This gives us a wave function $\hat{\mathcal{P}}$.
The entanglement entropy of $\hat{\mathcal{P}}$ can be computed from the reduced density matrix
\begin{equation}
\rho = {\hat{\mathcal{P}} \hat{\mathcal{P}}^\dagger \over \text{Tr} \hat{\mathcal{P}} \hat{\mathcal{P}}^\dagger}\ .
\end{equation} The Von Neumann entropy of $\rho$ is given by
\begin{equation}
S=-\text{Tr} \rho\log\rho\ .
\end{equation}
When $\ell=1$, the entropy vanishes only when $k$ is a multiplier of $N$ and the $k$ dependence is $\log{N\over \gcd(N,k)}$. Now fix $k=1$ and change $\ell$. It is interesting to find that the $\ell$ dependence of the entropy is $\log{N\over \gcd(N,\ell)}$. As observed in the $U(1)_k$ case, the $\ell$-linking is equivalent to linking $\ell$ times since we are only dealing with abelian symmetry lines. The $\ell$-times linking will decouple the subgroup $Z_{\gcd(N,\ell)}$. Alternatively if one treats the $\ell$-linking as the definition of a new symmetry transformation, the faithful group becomes order ${N\over \gcd(N,\ell)}$. Remarkably the $k$ dependence of the entropy and the $\ell$ dependence are exactly the same.

\subsection{SO$(2N+1)_k$}
For SO$(2N+1)$ Chern-Simons theory with level $k$, the 1-form symmetry is $Z_2$. One can use a set of non-negative integer, the Dynkin labels $(a_1,\cdots, a_{N})$, to label each integrable representation $R$. The integrable representations are constrained by $\phi_1 a_1+\cdots+\phi_{N}a_{N}\leq k$, where $k$ is the Chern-Simons level and $(\phi_1,\cdots, \phi_{N})$ are the comarks. For so$(2N+1)$ algebra this is given by $(1,2,\cdots,2,1)$. The modular S matrix is a matrix in the Hilbert space including all integrable representations. For two representations, $a=(a_1,\cdots, a_{N})$ and $b=(b_1,\cdots, b_{N})$, the modular S matrix is given by
\begin{equation}
\mathcal{S}_{a,b}=(-1)^{N(N-1)\over 2}{2^{N-1}\over (2N+k-1)^{N\over 2}}\text{Det}_{ij}\left[\sin\left({2\pi \phi_{a}[i]\phi_{b}[j]\over 2N+k-1}\right)\right]\ .
\end{equation} where $i,j =1,\cdots, N$. For a given representation, the function $\phi$ is
\begin{equation}
\phi_a[i] = L_i-i+{2N+1\over 2}\ ,
\end{equation} where $L_i=\sum_{m=i}^{N-1} a_m+{a_N\over 2}$ and $L_N={a_N\over 2}$. The modular T matrix is given by
\begin{equation}
\mathcal{T}_{a,b}=\delta_{a,b}\exp\left(-2\pi i {kN(2N+1)\over 24 (2N+k-1)}\right)\exp\left[{2\pi i\over 4(2N+k-1)}\left(2x_a+{N\over 2}y_a\right)\right]\ ,
\end{equation} where
$x_a$ and $y_a$ are integers determined by Dynkin labels
\begin{equation}
x_a = \sum_{i=1}^{N-1} a_i\left(\sum_{j=1}^i j a_j+\sum_{j=i+1}^{N-1}ia_j +(2Ni-i^2+ia_N)\right)\ ;\quad  y_a =a_N(a_N+2N)\ .
\end{equation}
Truncation means we only consider a $2\times 2$ matrix covering the $Z_2$ symmetry lines. We denote the truncated S matrix as $\hat{\mathcal{S}}$.
Similarly the truncated T matrix is a $2\times 2$ matrix with row and column corresponding to those of the truncated modular S matrix. We denote it as $\hat{\mathcal{T}}$.
We consider a two component torus link with linking number $\ell$ in the truncated Hilbert space. By $\hat{\mathcal{S}}$ and $\hat{\mathcal{T}}$ we obtain a wave function $\hat{\mathcal{P}}$.
The entanglement entropy of $\hat{\mathcal{P}}$ can be computed from the reduced density matrix. In this case the entropy vanishes for any integer $k$ and $\ell$.

\subsection{Sp$(2N)_k$}

For Sp$(2N)$ Chern-Simons theory with level $k$, the 1-form symmetry is $Z_2$. The integrable representations are constrained by $\phi_1 a_1+\cdots+\phi_{N}a_{N}\leq k$, where $k$ is the Chern-Simons level and $(\phi_1,\cdots, \phi_{N})$ are the comarks. For Sp$(2N)$ algebra this is given by $(1,1,\cdots,1,1)$. The modular S matrix is a matrix in the Hilbert space including all integrable representations. For two representations, $a=(a_1,\cdots, a_{N})$ and $b=(b_1,\cdots, b_{N})$, the modular S matrix is given by
\begin{equation}
\mathcal{S}_{a,b}=(-1)^{N(N-1)\over 2}{2^{N\over 2}\over (N+k+1)^{N\over 2}}\text{Det}_{ij}\left[\sin\left({\pi \phi_{a}[i]\phi_{b}[j]\over N+k+1}\right)\right]\ .
\end{equation} where $i,j =1,\cdots, N$. For a given representation, the function $\phi$ is
\begin{equation}
\phi_a[i] = L_i-i+N+1\ ,
\end{equation} where $L_i=\sum_{m=i}^{N} a_m$. The modular T matrix is given by
\begin{equation}\mathcal{T}_{a,b}=\delta_{a,b}\exp\left(-2\pi i {kN(2N+1)\over 24 (N+k+1)}\right)\exp\left[{\pi i\over 2(N+k+1)}x_a\right]\ ,
\end{equation} where
$x_a$ and $y_a$ are integers determined by Dynkin labels
\begin{equation}
x_a = \sum_{i=1}^{N} a_i\left(\sum_{j=1}^i j a_j+\sum_{j=i+1}^{N}ia_j +(2Ni-i^2+i)\right)\ .
\end{equation}
Truncation means we only consider a $2\times 2$ modular S matrix covering the $Z_2$ symmetry lines, $\hat{\mathcal{S}}$.
Similarly the truncated T matrix is a $2\times 2$ matrix with row and column corresponding to those of the truncated modular S matrix. We denote it as $\hat{\mathcal{T}}$.
Fix $Nk$ to be odd, the entanglement entropy in this case is $\log 2$ when $\ell$ is odd and vanishes when $\ell$ is even.

\subsection{SO$(2N)_k$}
For SO$(2N)$ Chern-Simons theory with level $k$, the 1-form symmetry is $Z_2\times Z_2$ when $N$ is even and $Z_4$ when $N$ is odd. The integrable representations are constrained by $\phi_1 a_1+\cdots+\phi_{N}a_{N}\leq k$, where $k$ is the Chern-Simons level and $(\phi_1,\cdots, \phi_{N})$ are the comarks. For so$(2N)$ algebra this is given by $(1,2,\cdots,2,1,1)$. The modular S matrix is a matrix in the Hilbert space including all integrable representations. For two representations, $a=(a_1,\cdots, a_{N})$ and $b=(b_1,\cdots, b_{N})$, the modular S matrix is given by
\begin{equation}
\mathcal{S}_{a,b}=(-1)^{N(N-1)\over 2}{2^{N-2}\over (2N+k-2)^{N\over 2}}\left(\text{Det}_{ij}[M_{a,b}]+i^N \text{Det}_{ij}[G_{a,b}]\right)\ ,\end{equation} where $M_{a,b}$ and $G_{a,b}$ are $N\times N$ matrices whose elements are defined as
\begin{equation}
M_{a,b}[i,j]=\cos\left(2\pi\phi_a[i]\phi_b[j]\over 2N+k-2\right)\ ;\quad G_{a,b}[i,j]=\sin\left(2\pi\phi_a[i]\phi_b[j]\over 2N+k-2\right)\ .
\end{equation}
For a given representation, the function $\phi$ is
\begin{equation}
\phi_a[i] = L_i-i+N\ ,
\end{equation} where $L_i=\sum_{m=i}^{N-2} a_m+{a_N+a_{N-1}\over 2}$ and $L_{N-1}={a_N+a_{N-1}\over 2}$, $L_{N}={a_N-a_{N-1}\over 2}$. The modular T matrix is given by
\begin{equation}
\mathcal{T}_{a,b}=\delta_{a,b}\exp\left(-2\pi i {kN(2N-1)\over 24 (2N+k-2)}\right)\exp\left[{2\pi i\over 2(2N+k-2)}\left(x_a+{N\over 4}y_a-a_Na_{N-1}\right)\right]\ ,
\end{equation} where
$x_a$ and $y_a$ are integers determined by Dynkin labels
\begin{equation}
x_a = \sum_{i=1}^{N-2} a_i\left(\sum_{j=1}^i j a_j+\sum_{j=i+1}^{N-2}ia_j +i(2N-i-3)\right)+(a_{N-1}+a_N+2)\sum_{j=1}^{N-2}ja_j\ ;
\end{equation}
\begin{equation}
y_a =(a_N+a_{N-1})^2+2(N-1)(a_N+a_{N-1})\ .
\end{equation}
Truncation means we only consider a $4\times 4$ matrix covering the $4$ symmetry lines. We denote the truncated S matrix as $\hat{\mathcal{S}}$.
Similarly the truncated T matrix is a $4\times 4$ matrix with row and column corresponding to those of the truncated modular S matrix. We denote it as $\hat{\mathcal{T}}$.
The wave function of a two component link with linking number $\ell$ in the truncated Hilbert space is given by
\begin{equation}
\hat{\mathcal{P}}_{j_1,j_2} = \sum_k (\hat{\mathcal{S}}\hat{\mathcal{T}}^\ell \hat{\mathcal{S}})_{0k}{\hat{\mathcal{S}}_{j_1k}\hat{\mathcal{S}}_{j_2k}\over \hat{\mathcal{S}}_{0k}}\ .
\end{equation}
The entanglement entropy can be computed from the reduced density matrix. When $N$ is even, the 1-form symmetry is $Z_2\times Z_2$. Fix $k$ to be odd, the entropy vanishes when $\ell$ is even and becomes $\log 4$ when $\ell$ is odd. This is because twice linking makes every element in $Z_2\times Z_2$ act twice and become identity. When $N$ is odd, the symmetry is $Z_4$. Fix $k$ to satisfy $\gcd(k,4)=1$, the $\ell$ dependence of entanglement entropy is $\log{4\over \gcd(4,\ell)}$.


\end{document}